\documentclass[sigconf, authorversion]{acmart}
\usepackage{multirow}
\makeatletter
\providecommand{\bibinfo}[2]{#2}
\providecommand{\bibfield}[2]{#2}

\providecommand{\showeprint}[1]{#1}
\providecommand{\showURL}[1]{#1}

\makeatother
\AtBeginDocument{%
  }

\setcopyright{none}

\begin{document}

\title{xSense Design Cards: Guiding the Design of Multisensory Experiences }


\author{Ceylan Beşevli}
\affiliation{%
\institution{Department of Computer Science, University College London}
\streetaddress{169 Euston Road}
\city{London}
\country{United Kingdom}
}

\author{Carlos Velasco}
\affiliation{%
\institution{Department of Marketing, BI Norwegian Business School}
\streetaddress{Nydalsveien 37}
\city{Oslo}
\country{Norway}
}

\author{Marianna Obrist}
\affiliation{%
\institution{Department of Computer Science, University College London}
\streetaddress{169 Euston Road}
\city{London}
\country{United Kingdom}
}

\renewcommand{\shortauthors}{Besevli et al.}
\settopmatter{printacmref=false}
\acmConference[]{}{}{}
\begin{abstract}
Designing multisensory experiences involves the deliberate combination of sensory elements to shape specific impressions for a given audience. Advances in technologies beyond audiovisual modalities now make it feasible to design across touch, taste, smell, and more. However, HCI still lacks the tools and shared vocabulary needed to systematically create and evaluate such experiences. The xSense Design Cards address this gap with four card types: (1) Experience Cards define purpose, context, and audience; (2) Sensory Cards break down multisensory concepts into elements and events; (3) Technology Cards prompt consideration of relevant technologies; and (4) Exploration Cards guide reflection on the broader context, including responsible innovation. This work introduces the cards and their theoretical grounding, showing how they support structured design, reflection, and evaluation of an experience’s multisensory composition. By presenting xSense, we aim to broaden the vocabulary for multisensory design and stimulate discussion within the growing multisensory HCI community.
   
\end{abstract}




\begin{teaserfigure}
  \includegraphics[width=\textwidth]{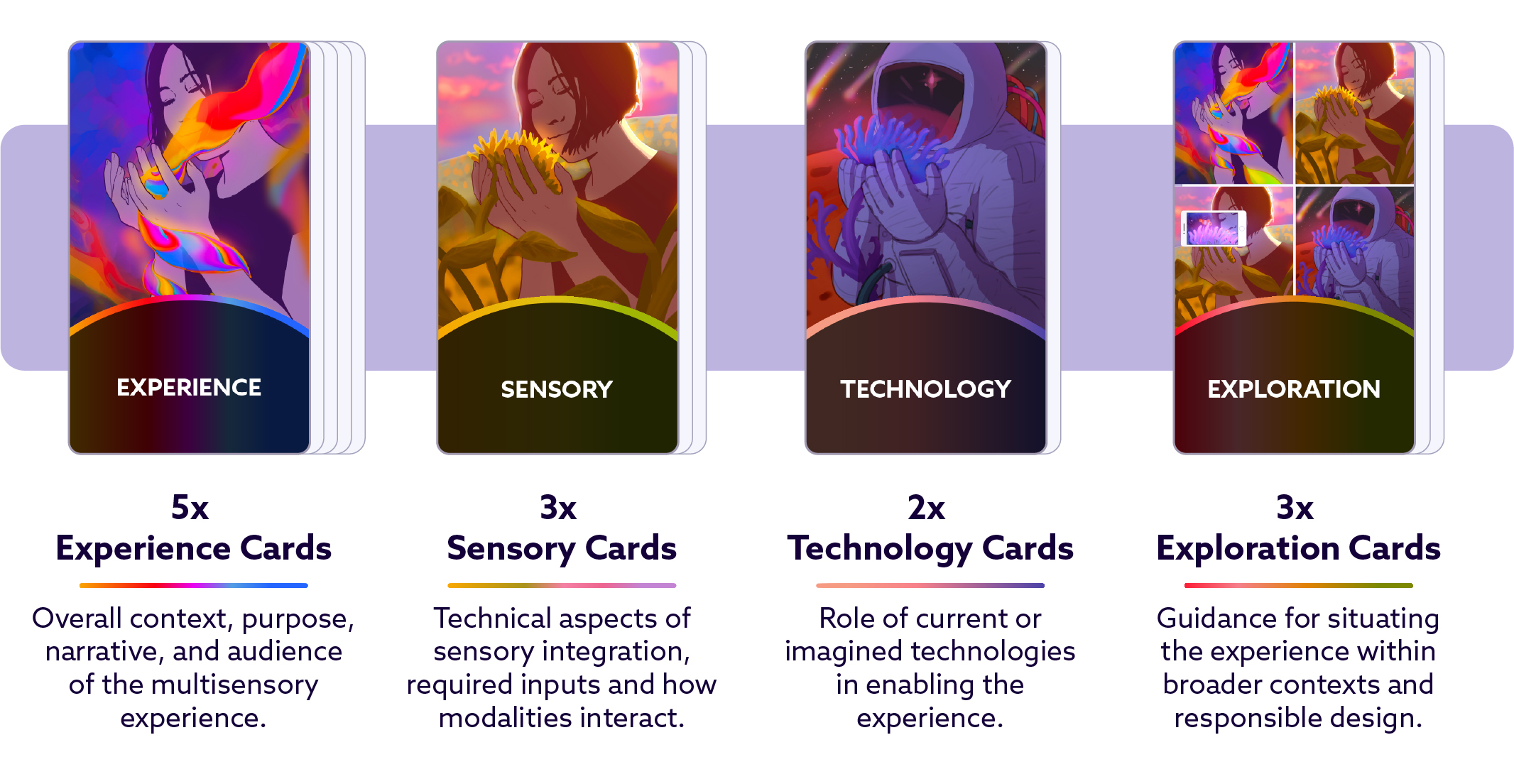}
  \caption{Overview of the xSense Design Cards categories}
  \Description{xSense Design Cards categories and explanations}
  \label{fig:teaser}
\end{teaserfigure}


\maketitle

\begin{figure*}
    \centering
    \includegraphics[width=1\linewidth]{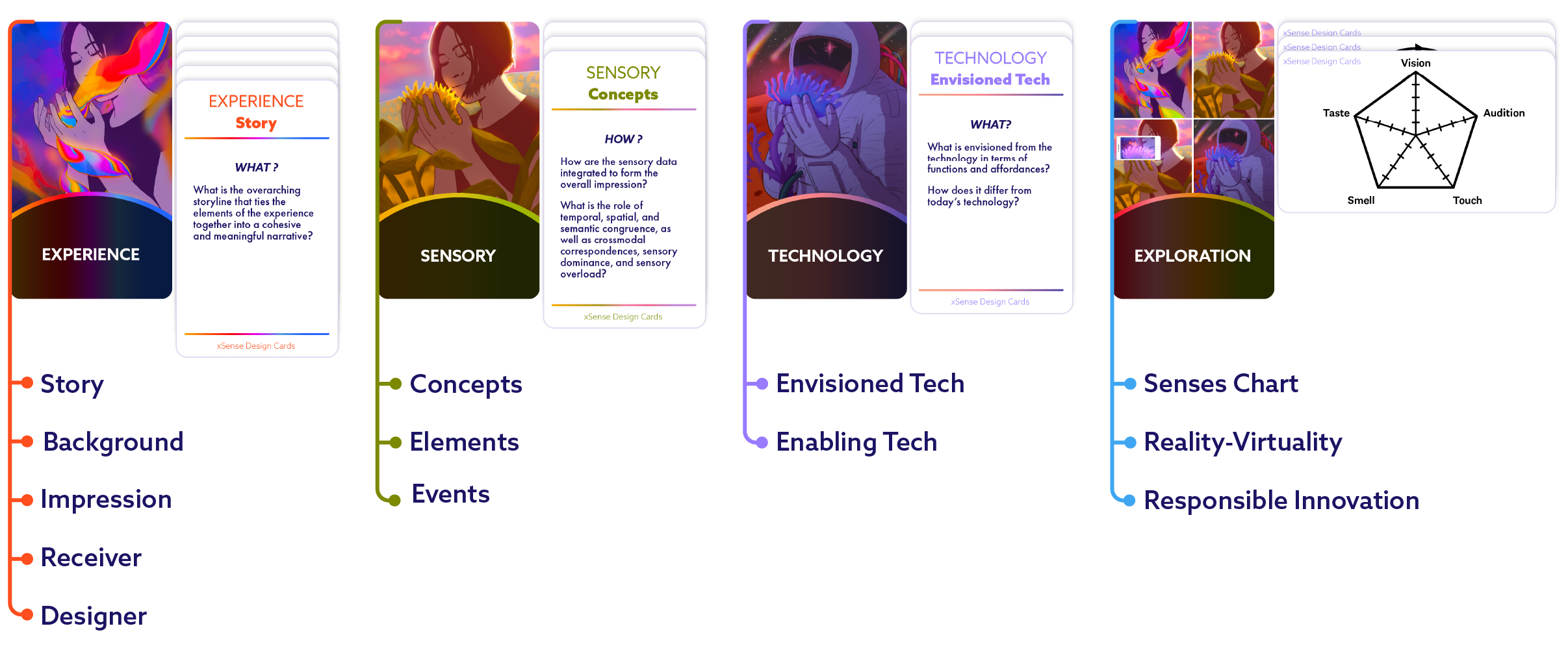}
    \caption{xSense Cards' contents with example cards from each category showing the prompt questions.}
    \label{fig:placeholder}
\end{figure*}

\section{Introduction}
Most everyday activities such as eating, commuting, socialising, or listening to music are inherently multisensory. Multisensory experiences are defined as "\textit{impressions formed by specific events whose sensory elements have been carefully crafted by someone}" \cite{velasco2025multisensory, velascoMultisensory1stEdition}.   
Advances in technologies for touch, taste, smell, and spatial audio are making it increasingly possible to design multisensory experiences intentionally and systematically \cite{cornelio2021multisensory, cornelio2023smell}. This momentum has sparked growing interest in multisensory interaction within Human–Computer Interaction (HCI) and related fields such as sensory marketing and cognitive psychology \cite{obrist2025multisensoryComm}.

Yet, despite the availability of domain-specific tools, such as flavour and aroma wheels (\cite{lee200whiskywheel, noble1984winewheel}), or tactile coding systems \cite{Fabtouch}, designing multisensory experiences remains challenging. Existing tools are often focused on individual senses or specific modalities, and rarely provide support for structuring impressions, mapping experience journeys, integrating multiple senses, or guiding storytelling across modalities \cite{velasco2025multisensory, velascoMultisensory1stEdition}. As a result, multisensory experiences are frequently developed through intuition rather than a systematic framework.

Design cards have long been valued in HCI for being simple, tangible, and easy to manipulate, helping teams structure discussion, explore perspectives, and establish a common language \cite{HsiehCards2023, roy2019card}. Building on these strengths, and proposal of \cite{velasco2025multisensory, velascoMultisensory1stEdition}, we introduce the xSense Design Cards, which is a deck developed to support the structured design, reflection, and evaluation of multisensory experiences. The deck draws on multisensory perception research and is organised into four complementary card types: \textit{Experience Cards} help designers define context and audience, \textit{Sensory Cards} guide the selection and combination of sensory elements, \textit{Technology Cards} consider current and future enabling tools, and \textit{Exploration Cards} support reflection on the experience’s position within conceptual spaces such as the reality–virtuality continuum \cite{milgram1995augmented,abowd2000charting}, and encourage ethical and responsible design practices \cite{stilgoe2020developing}.

By translating key principles of multisensory perception into a tangible, structured format, xSense Cards support the deliberate design, reflection, and evaluation of multisensory experiences. This paper presents the deck, outlines its conceptual grounding, and invites conversation and feedback to support further development of tools for multisensory design within the broader research and design communities.

\section{Background}

Designing multisensory experiences in HCI draws on both theoretical understanding of human perception and practical methods for guiding design. This section provides an overview of key components of multisensory experiences before introducing design cards as a method for structuring design processes and translating specialised knowledge into actionable prompts.

\subsection{Multisensory Experiences in HCI}

To design multisensory experiences systematically, it is useful to consider their conceptual components and how technology mediates perception. Following previous work \cite{velasco2025multisensory, obrist2025multisensoryComm}, we can define the components of multisensory experiences. First, the impression is the perceptual, emotional, and cognitive outcome arising from sensory stimulation, the lasting effect or understanding that results from the experience. Second, events are the temporal encounters that unfold over time, which can be viewed as single occasions or broken down into smaller moments (pre, during, post), often forming narratives through sequences. Third, sensory elements are the raw sensory data presented across the senses, ranging from low-level features (brightness, pitch) to high-level interpretations (faces, music). These elements integrate through concepts like spatio-temporal congruence \cite{spence2025merging}, semantic alignment \cite{laurienti2004semantic}, and crossmodal correspondences \cite{motoki2023crossmodal} to facilitate cohesive multisensory perception. Fourth, the receiver represents the audience with their unique characteristics, cultural background, demographics, personal history, personality traits, and psychographic profiles, all of which shape how the experience is perceived.
Technology enables multisensory experiences across the reality-virtuality continuum through specialised devices and interfaces. For example, virtual reality can manipulate lighting and 3D food shapes to enhance taste perception, AR can enrich walks with detailed sensory information, and 4DX cinema integrates touch and smell with careful temporal synchrony.

A brief survey of publication trends illustrates the growing interest in multisensory experiences: Google Scholar returns 663 results for the period 2000–2010 compared to 9,480 results from 2011–2025, indicating substantial growth in research activity. While a comprehensive review is beyond the scope of this paper, this increase highlights the need for tools that help designers systematically create and evaluate multisensory experiences.

\subsection{Design Cards}
Design cards have become a well-established format within HCI and design practice, valued for their simplicity, tangibility, and ease of use \cite{HsiehCards2023}. They provide an accessible way to introduce prompts into the design process, helping structure discussions and establish a shared vocabulary and understanding \cite{DengTangoCards}. At the same time, studies highlight several challenges: cards can overwhelm users with information or oversimplify complex concepts \cite{roy2019card}, may be difficult to apply without clear guidance, and are often hard to update once produced \cite{HsiehCards2023}. Overall, though, the effectiveness of design cards lies in their ability to provide structure to design processes

Design cards span a wide range of purposes and formats, with domain-specific decks representing nearly a quarter of all published sets \cite{roy2019card}. These cards translate specialised knowledge into accessible prompts, whether developed with experts \cite{BekkerDSD, BaykalCCICards}, grounded in theoretical or experiential frameworks \cite{PLEX}, or integrating user insights with disciplinary perspectives \cite{BalkayaHFICards}. xSense aligns with this domain-specific card category, offering a framework-informed approach to multisensory experience design.

\section{xSense Design Cards}  

The primary goal of xSense Design Cards is to provide a structured tool for guiding, analysing, and reflecting on the design of multisensory experiences. The complete deck consists of 13 cards organised into four categories: Experience (5 cards), Sensory (3 cards), Technology (2 cards), and Exploration (3 cards) (Figure 2). Table 1 provides an in-depth explanation of the card categories, their guiding questions and descriptions. 

The cards are designed for physical use but can also be adapted for digital formats, with print-ready PDFs and single-image versions available.

\begin{table*}[t]
\centering
\small
\begin{tabular}{p{1.7cm} p{1.9cm} p{1.5cm} p{8.5cm}}
\toprule
\textbf{Category} & \textbf{Card} & \textbf{Guiding Question} & \textbf{Description} \\
\midrule

\multirow{2}{*}{\textbf{Experience }}
& Background & \textit{Why?} & Specifies the experience by articulating its context and underlying reason. \\
& Impression & \textit{What?} & Frames the intended impression of an experience by specifying its perceptual, emotional, and cognitive outcomes. \\
& Story & \textit{What?} & Describes the narrative structure that binds sensory elements into a coherent experience; storytelling is central to meaning-making \cite{VelascoStorytelling, polletta2011sociology}. \\
& Designer &\textit{Who?} & Identifies who is crafting the experience, reinforcing the agency of the “someone” in the multisensory definition. \\
& Receiver & \textit{Whom?} & Characterises the intended audience and their cultural, demographic, and personal backgrounds, which shape interpretation \cite{chuquichambi2024individual, majid2015cultural}. \\

\midrule

\multirow{2}{*}{\textbf{Sensory }} 
& Events & \textit{When?} & Identifies the encounter or sequence of encounters that structure the experience, establishing its temporal unfolding \cite{obrist2014temporal, petitmengin2006describing}. \\
& Elements &\textit{Which?} & Details the sensory data presented (e.g., colours, sounds, scents, tastes), encouraging designers to consider level (low vs.\ high) \cite{velasco201levels} and dimension (prothetic vs.\ metathetic) \cite{stevens1957psychophysical}. \\
& Concepts & \textit{How?} & Encourages reflection on how sensory data integrate to form the impression, referencing key principles: temporal, spatial \cite{spence2025merging}, and semantic congruence \cite{laurienti2004semantic}; crossmodal correspondences \cite{motoki2023crossmodal}; sensory dominance \cite{fenko2010sensorydominance}; sensory overload \cite{scheydt2017sensoryoverload}.  \\

\midrule

\multirow{2}{*}{\textbf{Technology }} 
& Enabling Tech & \textit{What?} & Focuses on the devices, interfaces, and systems used to deliver the experience, encouraging reflection on technological possibilities and requirements. \\
& Envisioned Tech & \textit{What?} & Encourages imagining future technologies, their functions, and how they differ from what is feasible today. \\

\midrule

\multirow{2}{*}{\textbf{Exploration }} 
& Senses Chart & --- & Provides a radar chart for mapping the contribution of the five traditional senses within the experience. \\
& Reality--Virtuality Continuum & --- & Offers a diagram for situating the experience across the real--virtual spectrum, including Augmented Reality (AR) and Augmented Virtuality (AV) \cite{raisamo2019human, milgram1995augmented, abowd2000charting}. \\
& Responsible 

Innovation & --- & Draws on a four-dimension framework (Anticipate, Reflect, Engage, Act) to support ethical and societal reflection across the design lifecycle \cite{stilgoe2020developing}. \\
\bottomrule
\end{tabular}
\caption{Overview of the xSense Design Cards categories, their guiding questions and descriptions.}
\label{tab:xsense-cards}
\end{table*}

\subsection{Iterations of the Cards}
The initial version of the deck consisted of nine cards across five categories, including an early Considerations category that contained only a single card based on the Responsible Innovation framework. At this stage, the cards offered only high-level classifications without prompts, and several components were minimally represented. For example, the Experience category included only Background and Impression.

In the subsequent iteration, three co-authors with complementary expertise in design, HCI, and psychology (each with 5–20 years of multisensory research experience) expanded and refined the deck. This involved three key changes: we introduced question prompts for all cards to provide clearer guidance as recommended in prior work \cite{HsiehCards2023, roy2019card}; we added new subcategories, particularly within the Experience cards, which increased the total number of cards from 9 to 13; and we dissolved the Considerations category, integrating its Responsible Innovation element into the Exploration cards for better coherence.

Through informal use at events and small workshops, we observed that, even with prompts, users needed clearer explanations of what each card represents. In response, the latest iteration incorporates the conceptual definitions on the front of each card, supporting quicker interpretation and more confident use. We also have three 'Use Case' cards, where we explain how to use xSense for Ideation, Evaluation and Reflection (see Section 3.2).

\subsection{Use Cases}
xSense is designed as a flexible and adaptable tool that can support various phases of a user-centred design process. It is especially valuable in the early stages for exploring ideas, and later for reflecting on and evaluating existing multisensory experiences. Below, we illustrate the three modes of use, which are also included in the deck as instruction cards.

\subsubsection{Using xSense for Ideation}
When applied to ideation, xSense supports early conceptual development without restricting creativity.

We recommend beginning with the Experience Cards, particularly Impression and Story, to establish the purpose and audience of the experience. Once the narrative and motivation take shape, the Sensory Cards help expand and refine the concept by prompting consideration of sensory elements, modality interactions, and relevant perceptual principles.

We suggest introducing Technology Cards only after the concept has taken shape. Starting with technology can prematurely constrain creativity, as designers may focus too early on feasibility rather than the intended vision or impression. By considering technology later, teams can first define the experience they want to create and then identify current feasible solutions (Enabling Tech) while exploring future possibilities (Envisioned Tech) or inventive workarounds. 

Finally, Exploration Cards can help position the idea within wider conceptual and ethical contexts, such as where it sits on the reality–virtuality spectrum or what responsible innovation considerations might apply. 

By breaking an experience into distinct aspects, the cards allow the deck to function as a design space schema \cite{designspace2016Biskjaer}. In practice, this means each card becomes a parameter a designer can adjust, such as changing the Story, swapping sensory Elements, or reconsidering which technologies to use. Seeing these parameters laid out could make it easier to compare options during the ideation process.

\subsubsection{Using xSense for Evaluation}
xSense can also be applied as a structured tool for assessing multisensory experiences, focusing on outcomes rather than design intent, such as whether it achieves its intended goals, engages the audience effectively, and integrates sensory elements coherently.

The Experience Cards help evaluate whether the narrative, context, and audience considerations align with the original design intent: \textit{Does the experience communicate its intended message? Is it meaningful and engaging for the target audience?} The Sensory Cards guide assessment of the multimodal composition: \textit{Are the senses balanced and well-integrated? Are temporal, spatial, and semantic relationships effective?} Technology Cards support evaluation of the tools and systems used: \textit{Do the chosen technologies enable the intended impressions? Are there limitations or opportunities introduced by the tools themselves?} Finally, Exploration Cards encourage assessment of broader dimensions such as positioning along the reality–virtuality spectrum and alignment with ethical or societal considerations with the responsible innovation lens. This process also allows for systematic identification of strengths, gaps, and opportunities for improvement in a multisensory experience, supporting iterative refinement or post-deployment assessment. 

\subsubsection{xSense for Reflection}

Beyond evaluating an experience’s outcomes, xSense can serve as a reflective tool for designers to critically examine their own assumptions, decisions, and biases during or after a project. While \textit{Evaluation} examines whether an experience achieves its intended goals and engages the audience effectively, \textit{Reflection} focuses on the choices made by the designer and the rationale behind them. 

By systematically revisiting each card category, designers might ask: \textit{Whose story are we really telling, the institution’s, the scientist’s, or the visitor’s?} (Experience Cards - Receiver, Story). \textit{Which sensory modalities did we prioritise, and why? What did we leave out?} (Exploration Cards-Senses Chart). \textit{Did our technology choices constrain or enable the Impression we sought?} (Technology Cards). In addition to examining these high-level choices, reflection can also probe cultural and perceptual assumptions, asking: \textit{Did the semantic congruence we sought to achieve feel universal, or did it assume particular cultural associations?} (Experience Cards - Receiver) \textit{Could visitors who process sensory information differently still access the intended meaning?} (Sensory Cards - Elements). This structured reflection can reveal unintended hierarchies, such as consistently privileging vision over other senses, or defaulting to certain crossmodal pairings without questioning their cultural specificity \cite{obrist2025multisensoryComm, majid2015cultural}.

\section{Discussion, Outlook \& Conclusion}
The xSense Design Cards aim to open up the multisensory design space by lowering barriers to engagement and providing a flexible framework that can support designers at different stages, from ideation to evaluation and reflection. The deck also encourages situating outcomes within broader contexts, such as responsible innovation.

Future work will evaluate the xSense Design Cards with both experts and non-experts to assess their clarity, usability, and the effectiveness of the guidance they provide. Participants will use the cards for multiple purposes (e.g., ideation, evaluation, and reflection), and the deck will be refined based on these insights. We also aim to explore additional content or card types to support later stages of the design process, such as prototyping—a well-documented gap in existing design card tools \cite{HsiehCards2023}. These studies will inform refinements that maintain ease of use while expanding the deck’s relevance across diverse multisensory design contexts.

Looking forward, we see xSense as a foundation for more deliberate and reflective multisensory design. By providing an approachable framework, the deck may make multisensory composition more accessible and foster a shared vocabulary for designers. We hope it can encourage experimentation and critical engagement among researchers and practitioners across HCI and multisensory experience design. Integrating the deck with speculative, design futuring approaches \cite{Futuring} could help designers explore and prototype alternative multisensory futures \cite{Sonarios}, making potential experiences more tangible while prompting critical reflection on social, ethical, and sensory implications.


\end{document}